\documentclass{aip-cp}

\usepackage[numbers]{natbib}
\usepackage{url}
\usepackage[utf8]{inputenc}

\begin{document}

\title{The Joint Physics Analysis Center Website}

\author[aff1]{Vincent Mathieu\corref{cor1}}
\eaddress[url]{http://www.indiana.edu/$\sim$jpac/index.html}
\affil[aff1]{Center for Exploration of Energy and Matter, Indiana University, Bloomington, IN 47403}
\corresp[cor1]{mathieuv@indiana.edu}

\maketitle

\begin{abstract}
The Joint Physics Analysis Center is a collaboration between theorists and experimentalists working in hadronic physics. In order to facilitate the exchange of information between the different actors in hadron spectroscopy, we created an interactive website. In this note, I summarize the first projects available on the website.
\end{abstract}

\section{INTRODUCTION}
With the 12 GeV upgrade of the Thomas Jefferson National Laboratory (JLab), a new hall was created and a new detector, GlueX \cite{Dudek:2012vr}, was build. GlueX's primary goal is the search of exotic mesons via real photo-production of a fixed hydrogen target. This program complements the current hadron spectroscopy program of the COMPASS detector \cite{Abbon:2007pq}. Both facilities uses high energy beam (photons at 9 GeV for GlueX and pions at 190 GeV for COMPASS) that fragments into multiple mesons.  Revealing the spectrum of short-lived resonances decaying into these mesons requires a robust amplitude analysis~\cite{Battaglieri:2014gca}. To this end, parametrization of the reactions have to be proposed and fitted to the data. To facilitate this collaborative efforts between theorists and experimentalists, the Join Physics Analysis Center (JPAC) was created. After two years of activities, several papers were published by the JPAC collaboration and its members. The topics covered include the photoproduction of one  \cite{Mathieu:2015eia} or two mesons~\cite{Shi:2014nea} at high energies, elastic scattering with pion \cite{Mathieu:2015gxa} and kaon~\cite{Fernandez-Ramirez:2015tfa,Fernandez-Ramirez:2015fbq} and the three-body decays of heavy~\cite{Szczepaniak:2014qca,Szczepaniak:2015eza} and light~\cite{Guo:2015zqa,Danilkin:2014cra} mesons. Many projects have led to deliverables (such as codes) and a platform, an interactive webpage~\cite{webpage}, was created to exchanges material among the community. In this proceeding I summarize the projects and their features available online via the JPAC website.

\section{NEUTRAL PION PHOTOPRODUCTION AT HIGH ENERGIES}
The reaction $\gamma p \to \pi^0 p$ at high energies is controlled by the exchange of Reggeons with negative charge conjugation. The dominant exchanges, having the larger intercept, are the vector-like Reggeons $\omega$ and $\rho$ having isospin $I=0$ and $I=1$ respectively. Sub-dominant exchanges are the axial-vector Reggeons $h$ and $b$, whose influences show in the beam asymmetry.  

\begin{figure}[htb!]
	\includegraphics[width=0.8\linewidth]{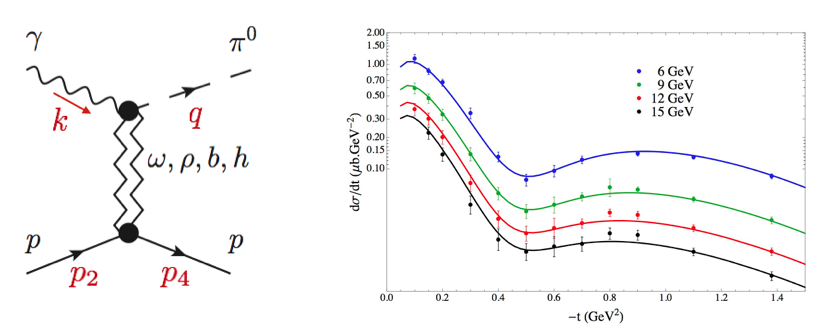}
        \caption {Left: Convention for the particle momenta. Right: Model compared to Anderson et al data \cite{Anderson:1971xh}.}
  \label{fig:mom}
\end{figure}

The parametrization of the helicity amplitudes are efficiently done using a the Chew-Goldberger-Low-Nambu (CGLN) scalar amplitudes~\cite{Chew:1957tf}. The amplitudes for the reaction $\gamma p \to \pi^0 p$ in its center-of-mass frame (the $s-$channel frame) is decomposed in the following way:
\begin{equation}
A^s_{\mu_4,\mu_2\mu_1}(s,t) = \bar u(p_4,\mu_4) \gamma_5
 F^{\mu\nu} (k,\lambda_1)
\left(
\frac{1}{2} \gamma_\mu\gamma_\nu\ A_1(s,t)
+ 2 q_\mu p_\nu\ A_2(s,t) 
+  q_\mu \gamma_\nu\ A_3(s,t) 
+  \frac{i}{2} \varepsilon_{\alpha\beta\mu\nu} \gamma^\alpha q^\beta\ A_4(s,t) \right) u(p_2,\mu_2).
\end{equation}
The photon field strength is $F^{\mu\nu} (k,\lambda_1) = \epsilon^\mu(k,\lambda_1) k^\nu -  k^\mu  \epsilon^\nu(k,\lambda_1)$ and the CGLN amplitudes are scalar function of the Mandelstam variable $s=(k+p_2)^2$ and $t=(k-q)^2$. The convention for the momentum are displayed on Figure~\ref{fig:mom}.

From the knowledge of the four scalar functions $A_i(s,t)$ all observables can be computed. The code is therefore organized in different blocks. The model is specified in a routine computing the scalar amplitudes. The helicity amplitudes in the $s-$channel are computed from the scalar amplitudes and the observables are calculated from the helicity amplitudes. The latter two functions are common to every parametrization and are inherent to the reaction. The user needs only to specify the scalar amplitudes to change the model. 

So far the model available online concerns the high energy domain $E_\gamma>4$ GeV where the dominance of Regge poles is manifest~\cite{Mathieu:2015eia}. The scalar amplitudes have good quantum numbers in the $t-$channel, the center-of-mass of crossed channel reaction $\gamma \pi^0 \to p \bar p$, cf. Table~\ref{tab:1}. It is then easy to parametrized the scalar amplitudes at high energies. The energy dependence and the phase are automatically predicted by the Regge theory and the remaining $t-$dependence of the residues are taken from one-particle-exchange model. 

\begin{table}
\caption{$\gamma \pi^0 \to p \bar p$ ($t-$channel)  quantum numbers of the CGLN scalar amplitudes.\label{tab:1}}
\label{tab:a}
\begin{tabular}{lccc}
\hline
  & \tch{1}{c}{b}{$J^{PC}$}  & \tch{1}{c}{b}{$I=0$}  & \tch{1}{c}{b}{$I=1$}    \\
\hline
$A_1$ & $\left(1,3,5,\ldots\right)^{--}$ & $\omega$ & $\rho$    \\
$A_2$ & $\left(1,3,5,\ldots\right)^{+-}$ & $h$ & $b$   \\
$A_3$ & $\left(2,4,6,\ldots\right)^{--}$ & $-$ &  $-$   \\
$A_4$ & $\left(1,3,5,\ldots\right)^{--}$ &  $\omega$ & $\rho$    \\
\hline
\end{tabular}
\end{table}

A Fortran code producing the differential cross section is available for download and for simulations on the website. In addition the interested user will find a $C$ code for the amplitudes and the AmpTools class associated. A Mathematica package, used to performed the fit, is also available.

\section{THREE-BODY DECAYS OF LIGHT MESONS}
\begin{figure}[htb!]
	\includegraphics[width=\linewidth]{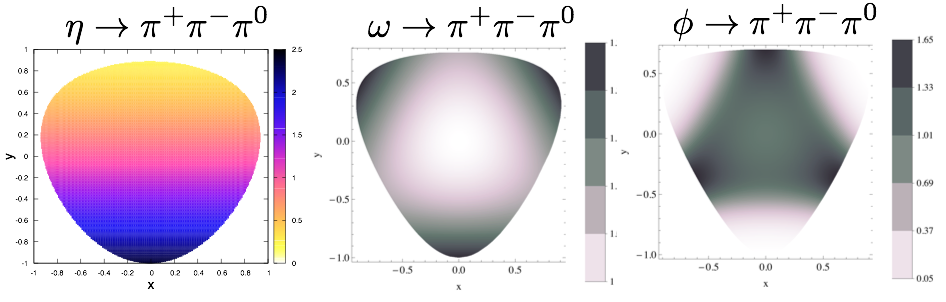}
        \caption {Dalitz plots in reduced variables for the three pions decays of eta, omega and phi mesons.}
  \label{fig:dalitz}
\end{figure}

\subsection{$\eta\to3\pi$}
From the theoretical point of view $\eta\to3\pi$ decays are of interest because of isospin violation. These decays are dominated by the intrinsic isospin breaking effects in Quantum Chromo-Dynamics (QCD) as electromagnetic effects are expected to be small. Consequently, the decay width for $\eta\to3\pi$ is expected to be proportional to the light quark mass difference and the decay amplitude is often expressed in terms of the quantity, $1/Q^2$ defined by
\begin{equation}
\frac{1}{Q^2} = \frac{m^2_d-m^2_u}{m_s^2-(m_d+m_u)^2/4}. 
\end{equation}
In the Reference~\cite{Guo:2015zqa}, we determined $Q$ by matching a parametrization of the Dalitz distribution $\eta\to \pi^+\pi^-\pi^0$ obtained by WASA@COSY~\cite{Bashkanov:2007aa} with the Next-to-Leading-Order (NLO) expansion of Chiral Perturbation Theory (ChPT)~\cite{Bijnens:2007pr}. 
Since the available phase space for this decay is small, the distribution will be described by only the lowest partial waves in each channels
\begin{equation}\label{eq:eta}
A(s,t,u) = \sum_{L=0}^{L_{{max}}=1} \frac{2L+1}{2} 
\left(\frac{2}{3} P_L(z_s) \left[a_{0L}(s)-a_{2L}(s)\right] +  P_L(z_t) \left[a_{1L}(t)+a_{2L}(t)\right] -P_L(z_u) \left[a_{1L}(u)-a_{2L}(u)\right] \right).
\end{equation}
The first index of the amplitudes $a_{IL}$ is the isospin, and $z_s$, $z_t$, $z_u$ are the cosine of the scattering angles in the $s$, $t$, $u$ channels respectively. Unitarity in the 3 channels imposes conditions on the amplitudes $a_{IL}$ written in the form a coupled channel integral equations for the functions $a_{IL}$. Indeed one sees easily that the partial waves, obtained by projection with $P_L(z_s)$ for the $L$ wave in the $s-$channel for instance, involves all the amplitudes $a_{IL}$. This is because of the non-orthogonality between Legendre polynomials with different arguments ($P_L(z_s)$ and $P_{L'}(z_t)$ for instance). The contribution from crossed-channels are called the ``3 body interactions". The model without the 3 body interactions is denoted the ``2 body" approximation and corresponds to the standard isobar model. As input we used two-pion scattering amplitudes from the analysis of~\cite{GarciaMartin:2011cn}. The parameters of the fit are the subtraction constants for each contributing partial wave, see the Reference~\cite{Guo:2015zqa} for the details. The Dalitz plot distribution fitted to the WASA@COSY data~\cite{Bashkanov:2007aa} is presented on Figure~\ref{fig:dalitz}.
The Dalitz plot is expressed using the dimensionless reduced variables $(x,y)$
\begin{equation}
x = \frac{\sqrt{3} (t-u)}{2 M(M-3\mu)}, \qquad \qquad \qquad
y = \frac{3 (M^2/3+\mu^2-s)}{2 M(M-3\mu)},
\end{equation}
with $M$ the mass of the decaying particles, the $\eta$ meson, and $\mu$ the pion mass.

The ChPT expression for the decay $\eta\to \pi^+\pi^-\pi^0$ is given by
\begin{equation}\label{eq:chpt}
A(s,t,u) = -\frac{1}{Q^2}\frac{m_K^2(m_K^2-m^2_\pi)}{3\sqrt{3}m^2_\pi F_\pi^2}
\left( M_0(s) - \frac{2}{3} M_2(s) + M_2(t)+ M_2(u) + (s-u)M_1(t) + (s-t)M_1(u)\right).
\end{equation}
The NNLO expressions of the isospin amplitudes $M_I$ can be found in~\cite{Bijnens:2007pr} and $F_\pi=92.3$ MeV is the pion decay constant. We matched Equation~(\ref{eq:eta}) and~(\ref{eq:chpt}) at the Adler zero $s=4/3m_\pi^2$ to determine the $Q$ value. We obtained
\begin{equation}
Q = 21.4\pm 0.4.
\end{equation}

The neutral decay $\eta\to \pi^0\pi^0\pi^0$ can be parametrized similarly. The codes for the charged and neutral modes of the $\eta$ decays are available for download and for simulations online. Both codes produces the amplitudes for the 2 body and 3 body models (with or without crossed-channel re-scattering) at given $(x,y)$ specified by the user.

\newpage
\subsection{$\omega,\phi\to3\pi$}
The next light meson decaying into three pions are the vector mesons. We continued our analysis of three pions with the $\omega$ and $\phi$ decays into  in Reference~\cite{Danilkin:2014cra}. In this case the spin of the vector meson is factorized using the Lorentz and isospin decomposition
\begin{equation}
A_\lambda^{abc}(s,t,u)  = i \varepsilon_{\mu\alpha\beta\gamma} \epsilon^\mu(p_V,\lambda) p_1^\alpha p_2^\beta p_3^\gamma \left( \frac{-i}{2} \epsilon^{abc} \right)  A(s,t,u).
\end{equation}
The scalar amplitudes $A(s,t,u)$ is then parametrized with a truncated partial wave expansion in all three channels as in Equations~(\ref{eq:eta}).  But  we kept only the $P-$wave since the higher waves $J=3,5,\ldots$ are expected to be insignifiant in the $\omega$ and $\phi$ decays into three pions:
\begin{equation}
A(s,t,u) = P'_1(z_s) F(s) + P'_1(z_t) F(t) +P'_1(z_u) F(u).
\end{equation}
The amplitude satisfy linear integral equations as in the $\eta$ decay
\begin{equation}\label{eq:omega}
\frac{1}{2i} \left[ F(s+i\epsilon) - F(s-i\epsilon)\right]  = \left(1-4\mu^2/s\right)^{1/2} t^*(s) \left( F(s) + \frac{3}{2} \int_{-1}^{1} (1-z_s^2) F(t[s,z_s])\, dz_s \right).
\end{equation}
In this unitarity equation, $t(s)$ represent the $P-$wave elastic $\pi\pi$ scattering. As in the $\eta$ decay we took the $\pi\pi$ phase-shift parametrization from the Reference~\cite{GarciaMartin:2011cn}. The strategies to solve Equation~(\ref{eq:omega}) are detailed in the Reference~\cite{Danilkin:2014cra}. We found that the 3 body effects, the second term in Equation~(\ref{eq:omega}), are negligible. The model predictions for the Dalitz distribution (normalized at the center of the Dalitz plot in the reduced variables) for the $\omega$ and $\phi$ are presented in Figure~\ref{fig:dalitz}. 

The Fortran code solving the integral equation and producing the amplitude for $\omega$ or $\phi$ decay in the 2 body or 3 body approximation are available for download and for simulations on the dedicated webpage. A Mathematica package is also available.

\section{COUPLED CHANNEL MODEL FOR $\bar K N$ SCATTERING}
\begin{figure}[htb!]
	\includegraphics[width=0.49\linewidth]{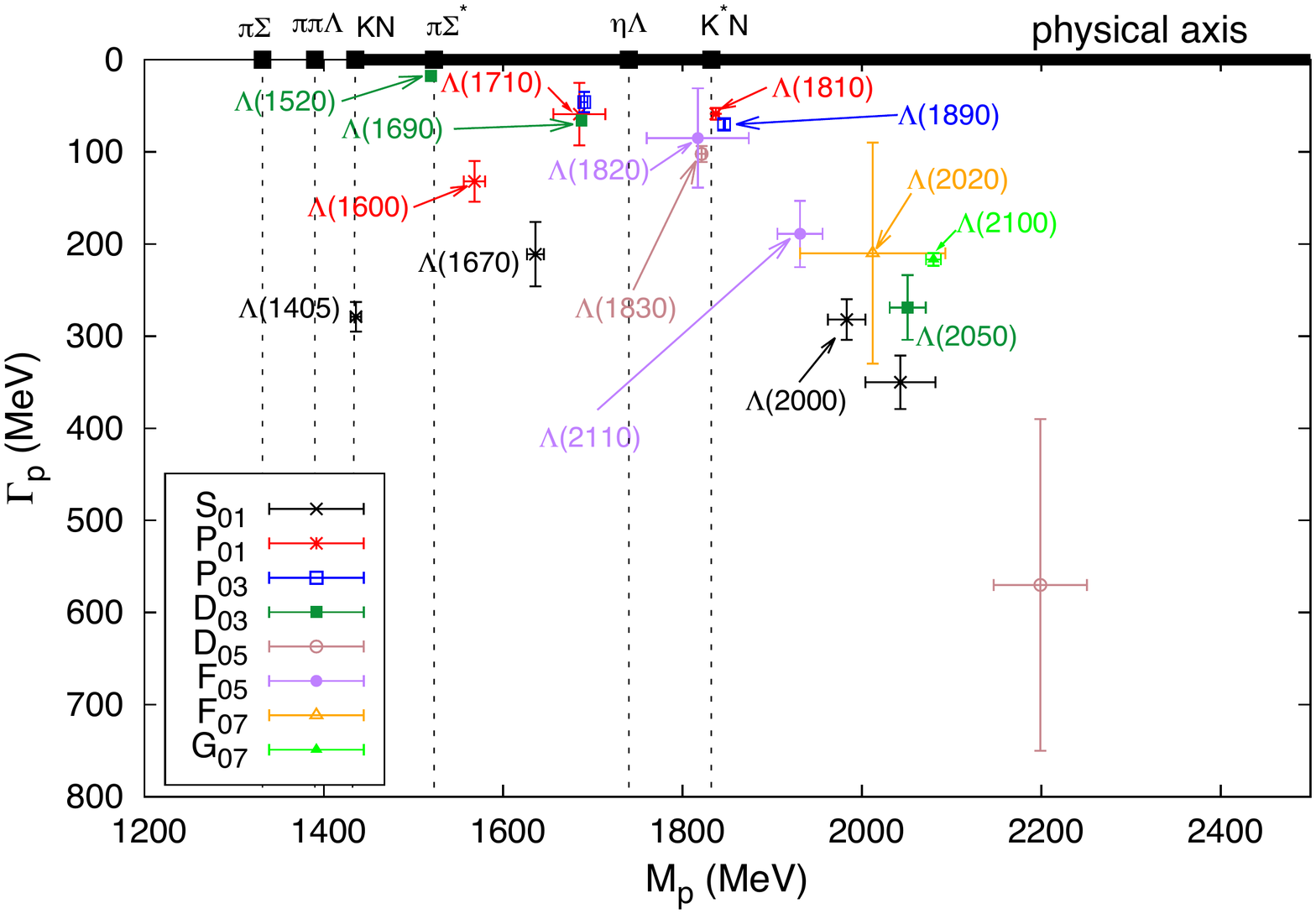}
	\includegraphics[width=0.49\linewidth]{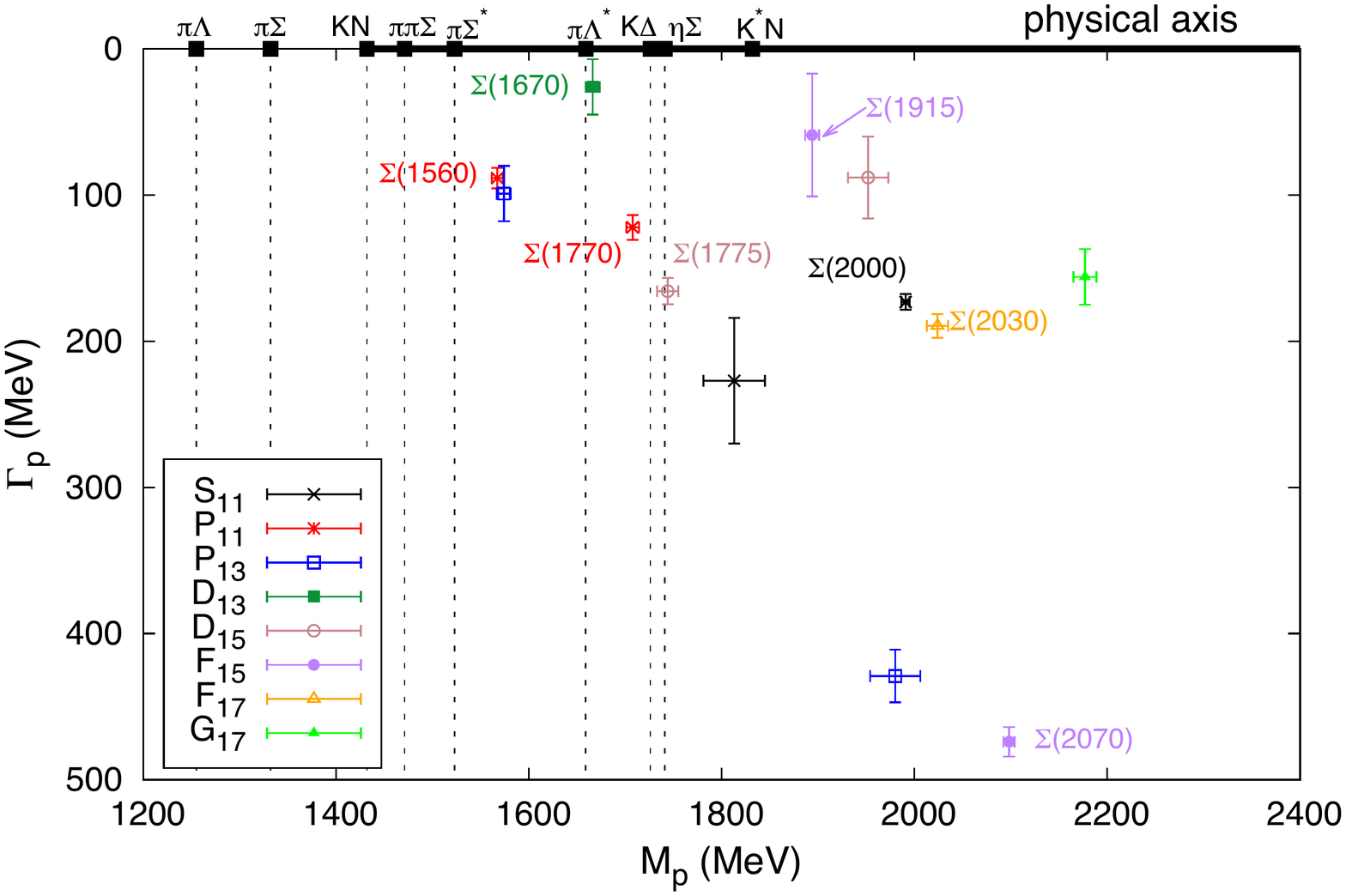}
        \caption {Spectrum of the $\Lambda$ (I=0) and $\Sigma$ (I=1) baryons from Reference~\cite{Fernandez-Ramirez:2015tfa}.}
  \label{fig:KN}
\end{figure}
In the Reference~\cite{Fernandez-Ramirez:2015tfa} we presented a unitary multichannel model for $\bar KN$ scattering in the resonance region that fulfills unitarity. Several coupled channels, indicated in the publication, were considered in the fitting procedure. In the JPAC webpage, 
the observables and partial waves for the following channels can be computed 
\begin{eqnarray}\nonumber
K^- p &\to& K^- p,\ \bar K^0 n,\\ \nonumber
&\to& \pi^- \Sigma^+,\ \pi^+ \Sigma^-,\ \pi^0 \Sigma^0, \\
&\to& \pi^0 \Lambda.
\end{eqnarray}
All observables, differential cross section $d\sigma/dz_s$, polarization observable $P$ and total cross section $\sigma$, are expressed in terms of the spin-non-flip $f(s,z_s)$ and the spin-flip $g(s,z_s)$ amplitudes with the relations
\begin{equation}
\frac{d\sigma}{dz_s}(s,z_s)  = \frac{1}{q^2} \left[ |f(s,z_s)|^2 + |g(s,z_s)|^2 \right], \quad
P\frac{d\sigma}{dz_s}(s,z_s)  = \frac{2}{q^2}  {Im}\left[ f(s,z_s)g^*(s,z_s) \right],  \quad
\sigma(s)  = \int_{-1}^1 \frac{d\sigma}{dz_s}(s,z_s)\ d z_s
\end{equation}
For a given channel (the channel index is omitted) the amplitudes admit a partial wave expansion
\begin{eqnarray}
f(s,z_s)  &=& \sum_{\ell=0}^{\infty} \left[ (\ell+1) R_{\ell+}(s) + \ell R_{\ell-}(s)\right] P_\ell(z_s), \\
g(s,z_s)  &=& \sum_{\ell=1}^{\infty} \left[ R_{\ell+}(s) - R_{\ell-}(s)\right] \sqrt{1-z_s^2}P'_\ell(z_s).
\end{eqnarray}
In a given meson-baryon channel $\ell$ labels the relative orbital angular momentum and the total angular momentum is given by $J = \ell \pm 1/2$. For a detailed relation, in all channels, between the orbital momentum and the partial waves we refer the reader to the Reference~\cite{Fernandez-Ramirez:2015tfa}.  

For a given orbital angular momentum $\ell$, we first remove the phase space factor with the introduction of a diagonal matrix $C_\ell(s)$
\begin{equation}
R_{\ell}(s) =  [C_\ell(s)]^{1/2} T_\ell(s) [C_\ell(s)]^{1/2}.
\end{equation}
Then the inverse of the reduced amplitudes satisfies a simple unitarity equations, Im $T^{-1}_\ell(s) = -i \rho(s,\ell)$, with $\rho(s,\ell)$ being the Chew-Mandelstam function. We can therefore use a real $K-$matrix to parametrize the reduced amplitudes as
\begin{equation}
T_\ell(s) = \left[ K^{-1}(s) - i \rho(s,\ell) \right]^{-1}.
\end{equation}
The $K-$matrix is the sum of the resonance contributions and a empirical background term. Each wave is parametrized and fitted independently. The detailed procedure is described in the Reference~\cite{Fernandez-Ramirez:2015tfa}. Finally the partial waves are analytically continue on the unphysical sheet and the pole positions are extracted. The resulting spectrum for $\Lambda$ and $\Sigma$ baryons is displayed on Figure~\ref{fig:KN}. 

The partial waves, binned in energy supplied by the user, can be dowloaded online . The Fortran code yielding the partial is also available. The differential cross section (together with the polarization) and the total cross section have also their dedicated pages. In the same spirit of all pages, the codes for producing the observables can be both simulated online and dowloaded.

\section{CONCLUSIONS}
After two years of activities, the JPAC has produced several papers concerning different hadronic reactions. The emphasis is given to the appropriate constraints related to the physics of the reaction. In the resonance region, one is interested extracting the properties of resonances (masses, widths and couplings) lying in the unphysical sheet. Therefore a particular care is given to unitarity, which controls the analytic continuation. In a three body decays of a light meson, resonances in different channels overlap. In that case, in addition to unitarity, crossing symmetry plays an important role and leads to integral equation between the amplitudes. In a high energy scattering, the number of relevant partial waves grows drastically and prevents the use of the partial wave expansion. However by an analytically continuation of the partial waves in the complex angular momenta plane, one can trade the partial expansion by an expansion in singularities in angular momentum, the Regge poles (and Regge cuts). The energy dependence and the phase are predicted and lead to simple parametrization of high energy data. 

The examples presented in this notes involve the extraction of physical quantities ($Q$ value, $\Lambda,\Sigma$ spectrum) and are buildings blocks for more complicated reactions. We hope that the codes available will help other physicists to describe other reactions such as the photoproduction of a pion, a $\eta$ or a $\omega$ meson. These processes are the basic reactions to be soon studies at the JLab facilities. We note also our $\bar K N$ amplitudes can be easily embedded in the photoproduction of a kaon pair (in which baryon resonances acts as a background for meson productions) but also in three body decays such as $\Lambda_b\to J/\psi K^- p$. 

The JPAC website is a part of broad collaborative project in hadron spectroscopy. We hope that the material available online will help the practitioners to quickly develop new parametrization for other reactions and update easily the current models.  As far as possible, we separated, in the codes and in the publication, the parametrization related to the kinematics (fixed for a given reaction) and the model-dependence. 

The website will grow as new projects are published. Previous project will be updated as new models or new codes are ready for sharing. We invite the community to browse the website regularly and send their comments to the JPAC members.

\section{ACKNOWLEDGMENTS}
This material is based upon work supported in part by the U.S. Department of Energy, Office of Science, Office of Nuclear Physics under contract DE-AC05-06OR23177. This work was also supported in part by the U.S. Department of Energy under Grant No. DE-FG0287ER40365, National Science Foundation under Grant PHY-1415459. 


\nocite{*}
\bibliographystyle{aipnum-cp}%

\begin{thebibliography}{99}

\bibitem{Dudek:2012vr} 
  J.~Dudek {\it et al.},
  Eur.\ Phys.\ J.\ A {\bf 48}, 187 (2012)
  [arXiv:1208.1244 [hep-ex]].


\bibitem{Abbon:2007pq} 
  P.~Abbon {\it et al.} [COMPASS Collaboration],
  Nucl.\ Instrum.\ Meth.\ A {\bf 577}, 455 (2007)
  [hep-ex/0703049].


\bibitem{Battaglieri:2014gca} 
  M.~Battaglieri {\it et al.},
  Acta Phys.\ Polon.\ B {\bf 46}, 257 (2015)
  [arXiv:1412.6393 [hep-ph]].


\bibitem{Mathieu:2015eia} 
  V.~Mathieu, G.~Fox and A.~P.~Szczepaniak,
  Phys.\ Rev.\ D {\bf 92}, no. 7, 074013 (2015)
  [arXiv:1505.02321 [hep-ph]].


\bibitem{Shi:2014nea} 
  M.~Shi, I.~V.~Danilkin, C.~Fern\'andez-Ramírez, V.~Mathieu, M.~R.~Pennington, D.~Schott and A.~P.~Szczepaniak,
  Phys.\ Rev.\ D {\bf 91}, no. 3, 034007 (2015)
  [arXiv:1411.6237 [hep-ph]].


\bibitem{Mathieu:2015gxa} 
  V.~Mathieu, I.~V.~Danilkin, C.~Fern\'andez-Ramírez, M.~R.~Pennington, D.~Schott, A.~P.~Szczepaniak and G.~Fox,
  Phys.\ Rev.\ D {\bf 92}, no. 7, 074004 (2015)
  [arXiv:1506.01764 [hep-ph]].


\bibitem{Fernandez-Ramirez:2015tfa} 
  C.~Fern\'andez-Ramírez, I.~V.~Danilkin, D.~M.~Manley, V.~Mathieu and A.~P.~Szczepaniak,
  arXiv:1510.07065 [hep-ph].


\bibitem{Fernandez-Ramirez:2015fbq} 
  C.~Fern\'andez-Ramírez, I.~V.~Danilkin, V.~Mathieu and A.~P.~Szczepaniak,
  arXiv:1512.03136 [hep-ph].


\bibitem{Szczepaniak:2014qca} 
  A.~P.~Szczepaniak and M.~R.~Pennington,
  Phys.\ Lett.\ B {\bf 737}, 283 (2014).


\bibitem{Szczepaniak:2015eza} 
  A.~P.~Szczepaniak,
  Phys.\ Lett.\ B {\bf 747}, 410 (2015)
  [arXiv:1501.01691 [hep-ph]].


\bibitem{Guo:2015zqa} 
  P.~Guo, I.~V.~Danilkin, D.~Schott, C.~Fern\'andez-Ramírez, V.~Mathieu and A.~P.~Szczepaniak,
  Phys.\ Rev.\ D {\bf 92}, no. 5, 054016 (2015)
  [arXiv:1505.01715 [hep-ph]].


\bibitem{Danilkin:2014cra} 
  I.~V.~Danilkin, C.~Fern\'andez-Ram\'irez, P.~Guo, V.~Mathieu, D.~Schott, M.~Shi and A.~P.~Szczepaniak,
  Phys.\ Rev.\ D {\bf 91}, no. 9, 094029 (2015)
  [arXiv:1409.7708 [hep-ph]].


\bibitem{webpage}
	\url{http://www.indiana.edu/~jpac/index.html}

\bibitem{Chew:1957tf} 
  G.~F.~Chew, M.~L.~Goldberger, F.~E.~Low and Y.~Nambu,
  Phys.\ Rev.\  {\bf 106}, 1345 (1957).


\bibitem{Anderson:1971xh} 
  R.~L.~Anderson, D.~Gustavson, J.~R.~Johnson, I.~Overman, D.~Ritson, B.~H.~Wiik and D.~Worcester,
  Phys.\ Rev.\ D {\bf 4}, 1937 (1971).


\bibitem{Bashkanov:2007aa} 
  M.~Bashkanov {\it et al.},
  Phys.\ Rev.\ C {\bf 76}, 048201 (2007)
  [arXiv:0708.2014 [nucl-ex]].


\bibitem{Bijnens:2007pr} 
  J.~Bijnens and K.~Ghorbani,
  JHEP {\bf 0711}, 030 (2007)
  [arXiv:0709.0230 [hep-ph]].


\bibitem{GarciaMartin:2011cn} 
  R.~Garcia-Martin, R.~Kaminski, J.~R.~Pelaez, J.~Ruiz de Elvira and F.~J.~Yndurain,
  Phys.\ Rev.\ D {\bf 83}, 074004 (2011)
[arXiv:1102.2183 [hep-ph]].

\end{thebibliography}

\end{document}